# Vortex Dynamics-Mediated Low-Field Magnetization Switching in an Exchange-Coupled System

Weinan Zhou,[1] Takeshi Seki,[1,2]* Hiroko Arai,[2,3]
Hiroshi Imamura,[3] Koki Takanashi[1]

[1]Institute for Materials Research, Tohoku University, Sendai, 980-8577, Japan
[2]JST PRESTO, Saitama, 332-0012, Japan
[3]National Institute of Advanced Industrial Science and Technology,
Tsukuba, 305-8568, Japan

*Corresponding author: go-sai@imr.tohoku.ac.jp

A magnetic vortex [1, 2] has attracted significant attention since it is a topologically stable magnetic structure in a soft magnetic nanodisk. Many studies have been devoted to understanding the nature of magnetic vortex in isolated systems. Here we show a new aspect of a magnetic vortex the dynamics of which strongly affects the magnetic structures of environment. We exploit a nanodot of an exchange-coupled bilayer with a soft magnetic $Ni_{81}Fe_{19}$ (permalloy; Py) having a magnetic vortex and a perpendicularly magnetized $L1_0$-FePt exhibiting a large switching field ($H_{sw}$). The vortex dynamics with azimuthal spin waves makes the excess energy accumulate in the Py, which triggers the reversed-domain nucleation in the $L1_0$-FePt at a low magnetic field. Our results shed light on the non-local mechanism of a reversed-domain nucleation, and provide with a route for efficient $H_{sw}$ reduction that is needed for ultralow-power spintronic devices [3].

(146 words)

A magnetic vortex in a soft magnetic disk is an in-plane curling magnetic structure having a core whose magnetic moments are normal to the disk plane [1,2]. Magnetic vortices have fascinated us because of their unique functionalities [4] and rich physics [5]. Several kinds of non-equilibrium dynamical motion can be excited by applying an rf magnetic field ($H_{rf}$) [5-7] or injecting spin current [8-10], leading to promising applications such as a vortex-type magnetic random access memory and a spin torque vortex oscillator. At certain conditions, the vortex polarity (core magnetization direction) and/or the circulation of in-plane magnetic moments can be switched [11-14]. Those studies focus on the control of magnetic moments in the vortex. Although interplay of the vortex in a magnet and the magnetization in an adjacent exchange-coupled magnet was investigated in a previous paper [15], no one has tried to use magnetic vortex dynamics in a soft magnet as a route to switching the magnetization of a hard magnet. Here, we show $H_{rf}$-induced vortex dynamics in soft magnetic Py non-locally triggers the magnetization switching of hard magnetic $L1_0$-FePt, which can balance competing goals for reducing $H_{sw}$ and maintaining the thermal stability of magnetization in a nanosized magnet.

We used the exchange-coupled system consisting of nanodots with a hard magnetic $L1_0$-FePt layer and a soft magnetic Py layer (**Figs. 1a and 1b**). The 10-nm-thick $L1_0$-FePt had large uniaxial magnetocrystalline anisotropy ($K$) along the perpendicular direction to the disk plane ($z$ direction), whereas the 150-nm-thick Py possessed negligible $K$, but the nanodot shape induced the shape magnetic anisotropy. This enabled us to saturate the magnetic moments of Py in the *z* direction by applying a dc magnetic field (*H*) perpendicular to the disk plane that was lower than the demagnetizing field for the thin film form **[16]**.

First, we performed micromagnetic simulations to reveal the equilibrium magnetic state. The simulated magnetization (*M*) versus *H* is shown in **Fig. 1c**, where *M* was normalized by the saturation value and *H* was applied along the *z* direction. The *M* - *H* curve exhibits a two-step behaviour. As *H* is swept from -10 kOe, *M* starts to increase at *H* ~ -2.5 kOe. As depicted in **Fig. 1d**, at *H* = 0 kOe, the magnetic vortex is formed in Py whereas all the magnetic moments in $L1_0$-FePt saturate along the -*z* direction. The vortex structure has a small deformation, in which the magnetic moments are slightly tilted from the azimuthal direction of the disk to the radial direction. As *H* increases to 2.6 kOe, the magnetic moments are tilted to the +*z* direction ($m_z$ in **Fig. 1e**), although the vortex structure is still maintained ($m_x$ in **Fig. 1e**). Increasing *H* to 5 kOe compresses the vortex structure in Py to the interface, in which the core polarity is switched (**Fig. 1f**). By comparing the cross-sectional *x* - *y* images near the interface (**Figs. 1e and 1f**), one can see that there is no remarkable change in the $m_x$ component even for the compressed vortex structure. These magnetic states are totally different from the spatially twisted magnetic structures observed in the in-plane magnetized $L1_0$-FePt | Py bilayers **[17]**.

Next, we experimentally examined the question of whether the vortex dynamics in Py affect $H_{sw}$ in $L1_0$-FePt. **Fig. 2a** displays the full $M$ - $H$ curve exhibiting two-step magnetization reversal behaviour similar to the simulation. When $H$ was swept from positive to negative, the magnetization switching of the $L1_0$-FePt occurred in the range from -6 kOe to -9 kOe. The minor magnetization curves showing spring-back behaviour also suggest that the magnetic moments of $L1_0$-FePt ($m^{FePt}$) switched in the hatched $H$ region in **Fig. 2a** (see **Supplementary Fig. S1**). In order to evaluate $H_{sw}$ of $L1_0$-FePt under the vortex dynamics excitation in Py, we measured the anisotropic magnetoresistance (AMR) effect for the nanodot array located on a coplanar waveguide (CPW) (see **Methods** and **Supplementary Fig. S2**). **Fig. 2b** shows the electrical resistance ($R$) as a function of $H$ without $H_{rf}$ being applied. At large positive $H$, *e.g.* $H$ = 9 kOe, all the magnetic moments in the bilayer were aligned with $H$, giving a low $R$ value. As $H$ decreased, $m^{FePt}$ maintained the positive value while the magnetic moments in Py ($m^{Py}$) rotated gradually, forming a spatially non-uniform magnetic structure. The part of $m^{Py}$ directed along the signal line of the CPW increased the value of $R$ due to the AMR effect. As $H$ decreased further, $m^{FePt}$ started to switch and eventually all the magnetic moments saturated again at $H$ = - 9 kOe. As indicated by the hatched areas in **Figs. 2a and 2b**, the $H$ regions showing the switching of $m^{FePt}$ in the $R$ - $H$ curve are in good agreement with those in the full $M$ - $H$ curve. In case of no $H_{rf}$, *i.e.* no excitation of vortex dynamics, $H_{sw}$ is obtained to be 8.6 kOe as indicated by the green arrows.

To excite the vortex dynamics in Py, we applied $H_{rf}$ transverse to the signal line of the CPW

by injecting rf power of 22 dBm, which corresponded to $H_{rf}$ = 200 Oe. The representative $\Delta R$ - $H$ curve is shown in **Fig. 2c**, where $\Delta R$ is the resistance change from $R$ at $H$ = -11 kOe. The frequency ($f$) of $H_{rf}$ was 11 GHz. The shape of $\Delta R$ - $H$ curve for $f$ = 11 GHz (solid circles) is rather different from that without $H_{rf}$ (open circles). $R$ sharply drops to the low $R$ state at $H$ = ±2.8 kOe, indicating that applying $H_{rf}$ with $f$ = 11 GHz significantly reduced $H_{sw}$. **Fig. 2d** summarizes $H_{sw}$ and $R$ as a function of $f$. Compared to the value of $H_{sw}$ with no $H_{rf}$ applied, one can see a small decrease in $H_{sw}$ in the whole $f$ region when 22 dBm of rf power was injected. This $f$-independent decrease is attributable to Joule heating caused by the high rf power injection. In addition to the $f$-independent decrease, a strong reduction of $H_{sw}$ is evident in the range of $11 \leq f \leq 17$ GHz. In this $f$ range, $H_{sw}$ gradually increases, and the values of $H_{sw}$ in $f \geq 18$ GHz are almost the same as those in 6 GHz $\leq f \leq$ 10 GHz. This reduction is not due to the Joule heating because $R$ reflecting the device temperature does not show any correlation with $H_{sw}$. We can also exclude the possibility that excitation of uniform magnetization dynamics in $L1_0$-FePt could lead to the low $H_{sw}$ because the resonance frequency of $L1_0$-FePt is estimated to be about 170 GHz. The numerical simulation reproduces the experimental results of $f$ dependence of $H_{sw}$ as shown by the open circles in **Fig. 2d**.

**Figs. 3a - 3e** show snapshots of the time evolution of $m_z$ for $f$ = 11 GHz and $H$ = 3.4 kOe (see also **Supplementary Movie 1**). The sliced planes of Py (top panels) and FePt (bottom panels) are 2 nm and 5 nm, respectively, away from the $L1_0$-FePt | Py interface. Inhomogeneous dynamics are

excited in the Py (**Figs. 3b - 3d**). At $t$ = 0.525 nsec, reversed-domain nucleation occurs in $L1_0$-FePt beneath the vortex core in Py (**Fig. 3c**). The reversed domain expands coherently in Py and $L1_0$-FePt (**Fig. 3d**). Similar inhomogeneous dynamics are excited for all conditions of $f$ when $H_{sw}$ is reduced (see **Supplementary Fig. S3**). Now let us discuss the magnetization switching process induced by vortex dynamics. Several excitations such as gyrotropic motion of vortex core, azimuthal spin waves, and radial spin waves have been observed in $H_{rf}$-induced vortex dynamics for a single soft magnetic disk [7, 18-20]. In order to assign which dynamical mode is responsible for the switching we observed, we calculated the deviation of the magnetization (d$m$) from the equilibrium state. The time evolution of the $z$-component d$m$ (d$m_z$) at $f$ = 11 GHz are displayed in **Fig. 4**. The borders between the red region (positive d$m_z$) and blue one (negative d$m_z$) correspond to nodes of spin waves. One sees that there are several nodes exhibiting clockwise rotation, in which the wave vectors of spin waves are along the azimuthal direction. In addition, we have the node surrounding the vortex core. This node is attributable to the standing spin wave along the radial direction. Consequently, we consider that the eigenmode is the azimuthal spin wave having the node in the radial direction. Near the vortex core located at the center, d$m_z$ shows the steep spatial change in the narrow region.

The azimuthal rotation of spin wave enlarges the area of vortex core or produces the multiple vortex cores in some cases (see **Supplementary Fig. S3**). When the area of vortex core reaches a critical size, the reversed-domain nucleation occurs in $L1_0$-FePt. We quantitatively evaluate the nucleation volume in $L1_0$-FePt from the measurement temperature ($T$) dependence of $H_{sw}$ without $H_{rf}$ (**Fig. 5a**). According to the Néel-Arrhenius law, the $T$ dependence of $H_{sw}$ can be fitted by [21]

$$H_{sw}(T) = H_{sw,0}(T)\left\{1 - \sqrt{\frac{k_B T}{E_0(T)}\ln\left(\frac{k_B T}{E_0(T)}H_{sw,0}(T)\frac{f_0}{R}\right)}\right\}, \tag{1}$$

where $H_{sw,0}$ is the switching field without thermal agitation, $k_B$ is the Boltzmann constant, and $E_0$ is the energy barrier given by the product of $K$ and the magnetic volume ($V$). $f_0$ is the attempt frequency ($10^9$ Hz). $R$ is the rate of $H$ sweep (10 Oe/sec). We assume no remarkable $T$ dependence of $H_{sw,0}$ as previously reported for FePt [22]. $H_{sw,0}$ and $E_0$ are obtained to be $14.4 \pm 0.7$ kOe and $(0.9 \pm 0.3) \times 10^{-18}$ J, respectively. Using $K = 3.1 \times 10^6$ J/m$^3$ evaluated experimentally from the $M$ - $H$ curves for the FePt single layer, we obtain $V = 300 \pm 100$ nm$^3$, which corresponds to the nucleation volume under the static $H$ ($V^{nuc, H}$). On the other hand, the nucleation volume under the excitation of vortex dynamics ($V^{nuc, D}$) is estimated numerically. Let us once again look at the case of FePt switching under $H_{rf}$ of f = 11 GHz and $H$ = 3.4 kOe. **Fig. 5b** displays simulated $m_z$ as a function of the position in the dot along the in-plane y direction ($D_y$) at various $t$. At $t$ = 0.51 nsec, the region having $m_z > 0.9$ appears, which is defined $V^{nuc, D}$. Thus $V^{nuc, D}$ is estimated to be ~ 800 nm$^3$. This $V^{nuc, D}$ is comparable to $V^{nuc, H}$, suggesting that both nucleation processes have comparable $E_0$. We calculated the $t$ dependence of total $m_z$ ($m_{z,total}$) and total energy ($E_{total}$), which is the sum of Zeeman energy ($E_Z$), demagnetizing energy ($E_d$), exchange energy ($E_{ex}$) and anisotropy energy ($E_{ani}$) (**Fig. 5c**). $E_{total}$ gradually increases just after starting the vortex dynamics excitation, and then $E_{total}$ decreases whereas $m_z$ increases. As shown in the middle panel of **Fig. 5c**, the time dependent energy difference ($\Delta E$) indicates that only $E_{ex}$ contributes to the increase of $E_{total}$. At $t$ = 0.51 nsec, the excess $\Delta E_{ex}$ = 5 × 10$^{-17}$ J is accumulated mainly in Py (bottom panel of **Fig. 5c**), which is large enough to overcome $E_0$ = (0.9 ± 0.3) × 10$^{-18}$ J. Consequently, the accumulated $E_{ex}$ in Py due to the vortex dynamics excitation becomes a non-local trigger for the reversed-domain nucleation in $L1_0$-FePt. Again we emphasize that mere vortex core switching does not nucleate the reversed domain in $L1_0$-FePt (**Fig. 1c**) and the vortex dynamics excitation is essential for the reversed domain nucleation. Our results lead to not only insight into the nucleation phenomena but also a new way for information writing of magnetic storage and spintronic applications using topologically unique magnetic structures.

**METHODS**

**Device Fabrication**

An ultrahigh vacuum magnetron sputtering system and an ion beam sputtering (IBS) system were used for thin film preparation. The thin films were grown on an MgO (100) single crystal substrate with the stack of MgO subs. || Fe (1) | Au (60) | FePt (10) | Py (150) | Au (5) | Pt (3) (in nanometer). All the layers except FePt were grown at ambient temperature. The FePt (001) layer was epitaxially grown on the Au (100) buffer layer using the magnetron sputtering system. The substrate temperature was set at 550 ºC. The high temperature deposition promoted the $L1_0$ ordering in FePt, resulting to a large $K$. As a result, the FePt layer had the easy magnetization axis normal to the film plane. After the FePt deposition, the sample was transferred to the IBS chamber to deposit the Py | Au | Pt layers. In order to induce the perpendicular component of magnetization in the soft magnetic Py at small $H$, the thin films were microfabricated into circular nanodots through the use of electron beam lithography and Ar ion milling to reduce the demagnetization field of Py. The scanning electron microscope (SEM) image shows that the dots are about 260 nm in diameter. The details concerning the thin film preparation and the microfabrication process are described in **[16]**.

**Device Characterization**

The magnetization curves for the dot arrays were measured using a superconducting quantum interference device magnetometer at room temperature. On the other hand, $R$ was measured for the dot array located on the CPW using a lock-in amplifier. The Au buffer layer was patterned into the CPW with the signal line of 4 μm × 50 μm, and an array of more than 1000 dots of the FePt | Py bilayer was placed on the signal line. A signal generator was connected to the device through the RF port of a bias-Tee to apply the rf power to the CPW, which generated $H_{rf}$ along the transverse direction to the signal line. $H$ was applied perpendicular to the sample. The AMR curves were averaged for three measurements, and a linear change in $R$ due to the resistance drift was calibrated.

**Micromagnetic simulation**

We have used the mumax$^3$ package [23] for full micromagnetic simulations of FePt | Py magnetic layers. The simulated structure was a cylindrical dot 200 nm in diameter and 10 nm in thickness for the FePt layer and 150 nm in thickness for the Py layer. The sample was divided into discrete computational cells, and the size of each cell was $3.125 \times 3.125 \times 2.5$ nm$^3$. The saturation magnetization and the uniaxial anisotropy constant for FePt were $M_s = 1.15 \times 10^6$ A/m and $K_u = 3.2 \times 10^6$ J/m$^3$, respectively. The easy axis of the FePt was in the $z$-direction as shown in **Fig. 1a**. The material parameters for Py were $M_s = 0.8 \times 10^6$ A/m, and $K_u = 0$ J/m$^3$. We chose the stiffness constant of $A = 1.3 \times 10^{-11}$ J/m for whole the system. In the $M$ - $H$ curve simulation, $H$ was applied along the $z$-direction from -10 kOe to 10 kOe and from 10 kOe and -10 kOe in 200 Oe stepwise increments. In this calculation, the damping parameter of $\alpha = 0.5$ was used for both materials in order to expedite relaxation to the equilibrium orientation. The magnetization dynamics under $H_{rf}$ were calculated by applying $\boldsymbol{H}_{ext}(t) = (0, H'\sin(2\pi ft), H)$, where $H'$ and $f$ are the amplitude and the frequency of $H_{rf}$, respectively. The amplitude $H'$ was set to be 200 Oe. For the magnetization dynamics calculation, the values of $\alpha$ were set at 0.1 for FePt and 0.01 for Py. Thermal effects were neglected throughout the simulation for simplicity.

**Acknowledgements**

This work was partially supported by a Grant-in-Aid for Young Scientists A (25709056) and Grant-in-Aid for Scientific Research S (23226001). The device fabrication was partly performed at Cooperative Research and Development Center for Advanced Materials, IMR, Tohoku University.

**Competing financial interests**

The authors declare that they have no competing financial interests.

**Author contributions**

T.S. and K.T. planned and supervised the study. W.Z. prepared the thin films, fabricated the devices, and carried out the measurements on the magnetic and electrical properties. W.Z. and T.S. analyzed the experimental data. H.A. and H.I. performed the micromagnetic simulation. All of the authors contributed to the physical understanding and manuscript preparation.

**Figure 1 | $L1_0$-FePt | permalloy (Py) exchange-coupled system**. **a,** Schematic illustration of a microfabricated dot. The coordinate axes are also shown. **b,** Scanning electron microscope image for the array of dots. **c,** Simulated result of magnetization (*M*) versus the dc magnetic field (*H*), where *M* was normalized by the saturation value and *H* was applied along the *z* direction. **d,** Simulated structures of magnetic moments (*m*) of the *x* component ($m_x$, upper panel) and the *z* component ($m_z$, lower panel) at *H* = 0 Oe. **e** and **f,** Simulated magnetic structures of $m_x$ in the *y* - *z* plane (left panel), $m_z$ in the *y* - *z* plane (middle panel), and $m_x$ in the *x* - *y* plane (right panel) at *H* = 2.6 kOe and 5 kOe.

**Figure 2 | Evaluation of switching field ($H_{sw}$) for microfabricated dots with $L1_0$-FePt | Py.**

**a,** $M$ as a function of $H$, where the value of $M$ was normalized by the saturation magnetization. **b,** Electrical resistance of device ($R$) as a function of $H$. The rf magnetic field ($H_{rf}$) was not applied. The red (blue) circles denote the results when $H$ was swept from positive (negative) to negative (positive). **c,** $\Delta R$ - $H$ curves with $H_{rf}$ (solid circles) and without $H_{rf}$ being applied (open circles). $\Delta R$ is the resistance change from $R$ at $H$ = -11 kOe. 22 dBm of rf power was applied to the device, which corresponded to $H_{rf}$ = 200 Oe. The frequency ($f$) of $H_{rf}$ was set at 11 GHz. For both $M$ - $H$ and $R$ - $H$ curves, $H$ was applied perpendicular to the device plane. The green arrows denote $H_{sw}$ of $L1_0$-FePt. **d,** $H_{sw}$ (top panel) and $R$ (bottom panel) as a function of $f$. The red dotted line denotes $H_{sw}$ without $H_{rf}$, and the solid (open) circles represent the experimental (simulated) results. $R$ is the value at $H$ = -11 kOe.

**Figure 3 | Time evolution of magnetic structures under excitation of vortex dynamics.**

**a - e,** Snapshots for $m_z$ at $H$ = 3.4 kOe under the application of $H_{rf}$ with $f$ = 11 GHz. Top and bottom panels are $x$ - $y$ plane images for Py and FePt, respectively. The planes of the Py and FePt slices are 2 nm and 5 nm, respectively, away from the FePt | Py interface.

**Figure 4 | Time evolution of deviated magnetic structures from the equilibrium state. a**

- **e,** Snapshots for deviation of $m_z$ (d$m_z$) for Py at $H$ = 3.4 kOe under the application of $H_{rf}$ with $f$ = 11 GHz. The plane of the Py slice was 2 nm away from the FePt | Py interface.

**Figure 5 | Volume and energy for reversed-domain nucleation. a,** $H_{sw}$ as a function of measurement temperature ($T$), which was experimentally obtained from the temperature dependence of $M$ - $H$ curves without $H_{rf}$ being applied. The red solid line denotes the result of fitting. **b,** $m_z$ as a function of the position in the dot along the in-plane y direction ($D_y$) at various times ($t$). $D_y$ is denoted in the inset. $H$ was set at 3.4 kOe. The plane of the FePt slice was 5 nm away from the FePt | Py interface. **c,** (Top panel) Calculated $t$ dependence of total $m_z$ ($m_{z,total}$) and total energy ($E_{total}$). Since $H$ was set at 3.4 kOe, $m_{z,total}$ shows the value of ~ 0.7 before the reversed-domain nucleation. (Middle panel) $t$ dependent energy difference ($\Delta E$) for Zeeman energy ($E_Z$), demagnetizing energy ($E_d$), exchange energy ($E_{ex}$) and anisotropy energy ($E_{ani}$). (Bottom panel) $t$ dependence of $E_{ex}$ in Py and $L1_0$-FePt.

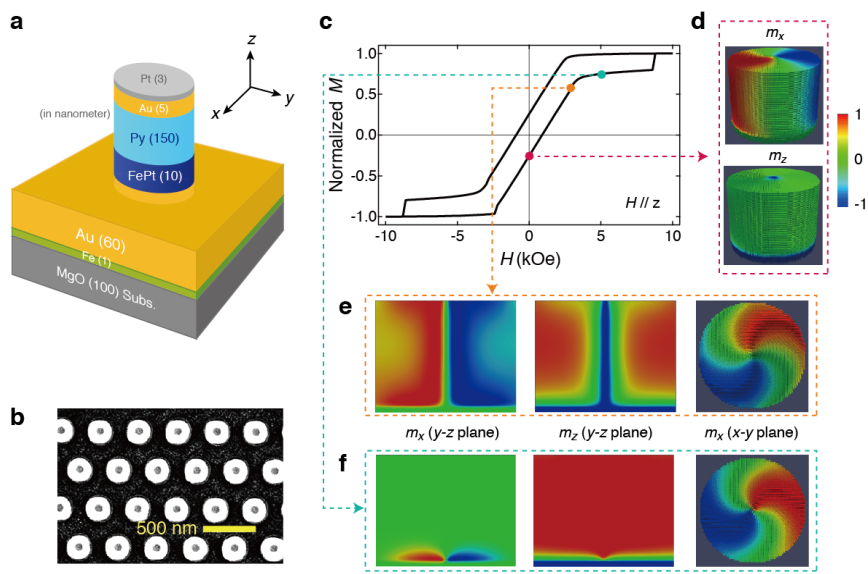

Figure 1

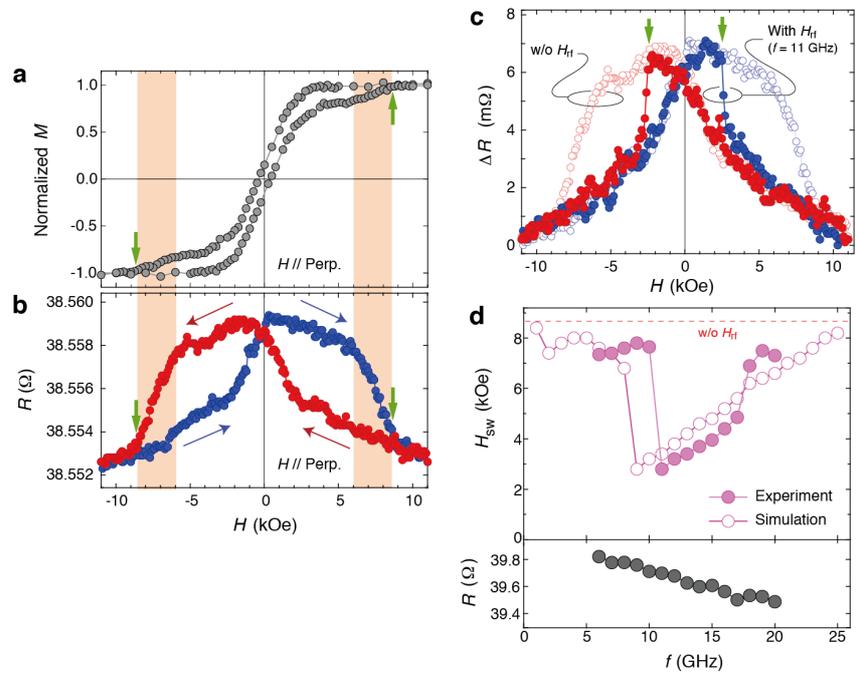

Figure 2

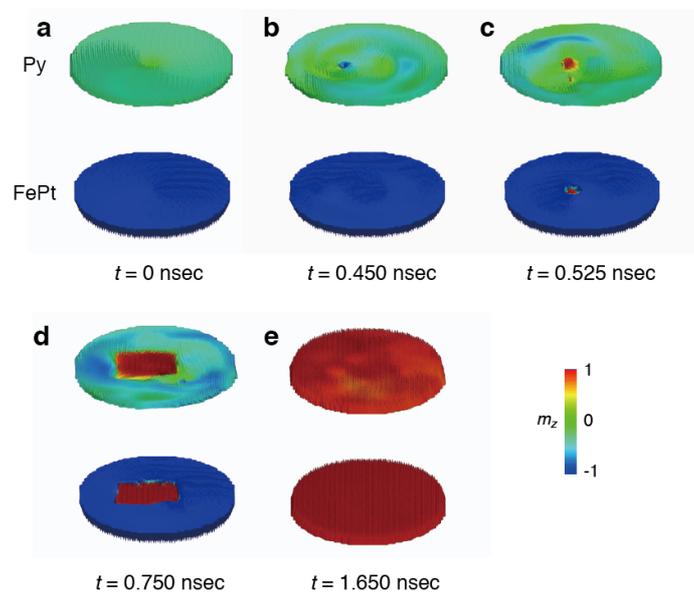

Figure 3
Zhou *et al.* Page 21

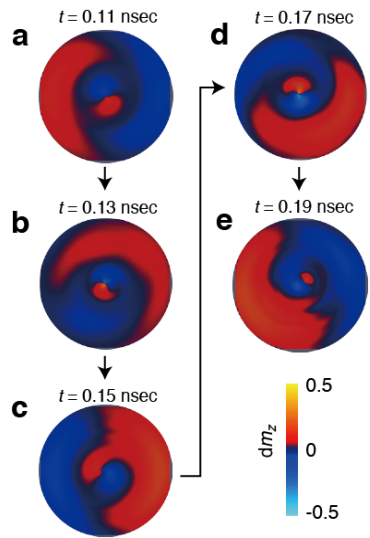

Figure 4

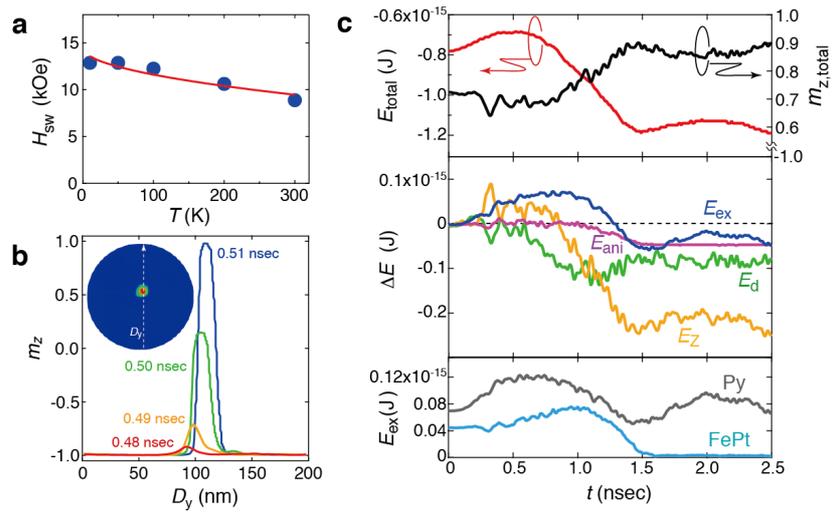

Figure 5

# Vortex Dynamics-Mediated Low Field Magnetization Switching in an Exchange-Coupled System
-Supplementary Information-

Weinan ZHOU, Takeshi SEKI, Hiroko ARAI, Hiroshi IMAMURA, Koki TAKANASHI

This Supplementary Information includes 3 figures (Figures S1 - S3).

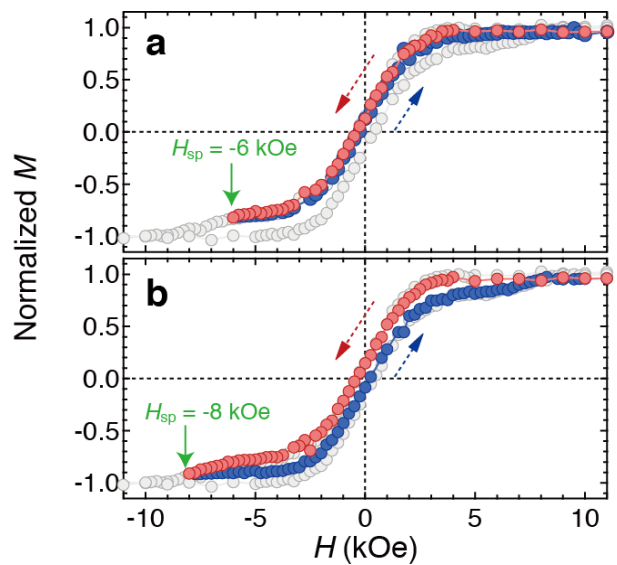

**Supplementary Figure S1 | Spring-back behavior in nanodots with exchange-coupled $L1_0$-FePt | Py. a,** Minor magnetization curve, where the perpendicular magnetic field ($H$) direction was reversed at $H_{sp}$ = -6 kOe. $H$ was swept from 10 kOe to -6 kOe, and then $H$ was swept to 10 kOe. **b,** Minor magnetization curve with $H_{sp}$ = -8 kOe. The gray solid circles denote the full magnetization curve. The reversible minor magnetization curve was obtained for $H_{sp}$ = -6 kOe whereas the hysteresis appeared for $H_{sp}$ = -8 kOe. This indicates that the magnetic moments of $L1_0$-FePt are switched in a part of dots when $H$ was swept to -8 kOe.

Zhou *et al.* Page 24

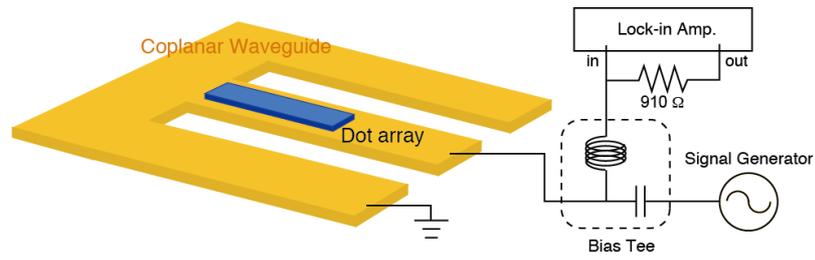

**Supplementary Figure S2 | Setup of electrical measurement for evaluating the switching field.** The electrical resistance of the device was measured for the dot array located on the coplanar waveguide (CPW) using a lock-in amplifier. A signal generator was connected with the device through the RF port of a bias-Tee to apply the rf power to the CPW, which generated the rf magnetic field along the transverse direction to the signal line. The magnetic field was applied perpendicular to the device plane.

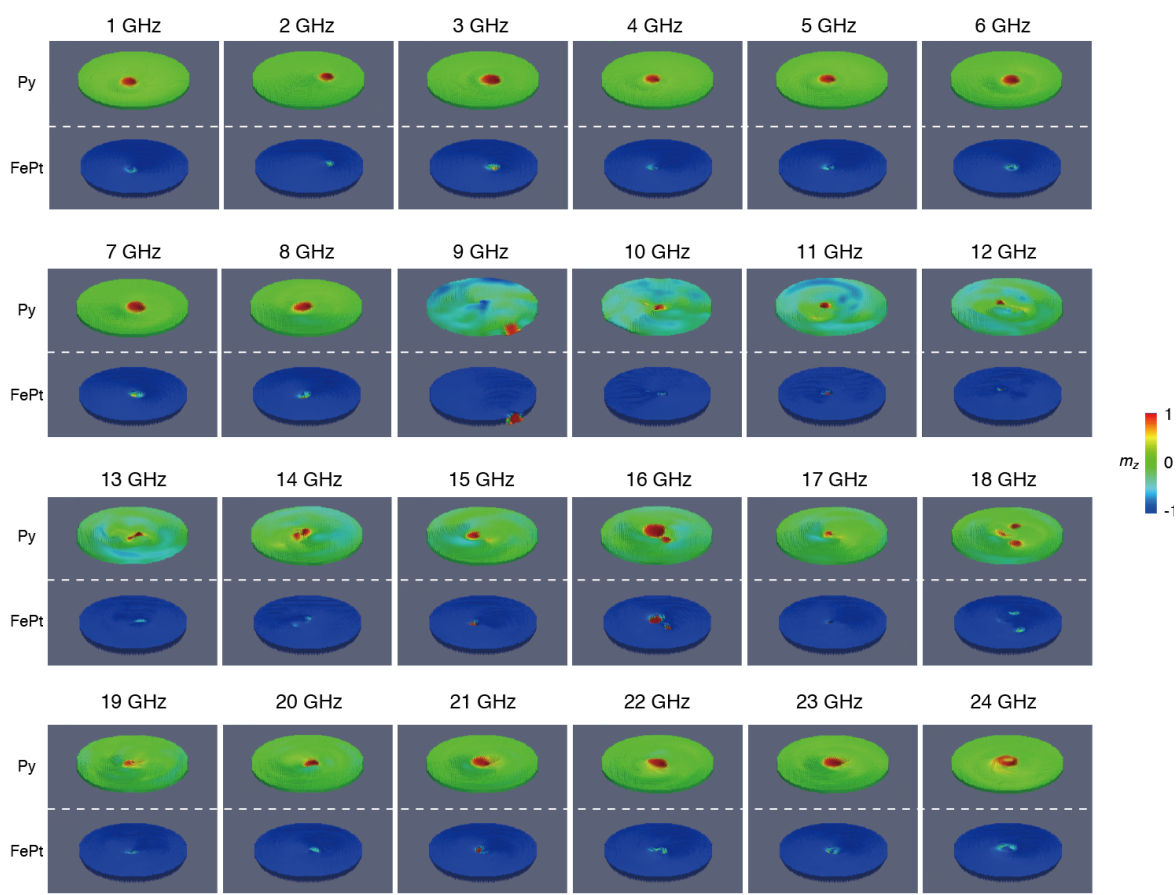

**Supplementary Figure S3 | Snapshots of simulated $m_z$ at the nucleation of switched domain in $L1_0$-FePt.** The frequency was varied in the range from 1 GHz to 24 GHz. Top and Bottom panels are the *x - y* plane images for Py and FePt, respectively. The sliced planes of Py and FePt are 2 nm and 5 nm away from the Py | FePt interface.